\documentclass[12pt]{iopart}
\usepackage{iopams}  
\usepackage[export]{adjustbox}

\expandafter\let\csname equation*\endcsname\relax
\expandafter\let\csname endequation*\endcsname\relax
\usepackage{mathtools}
\usepackage{amsmath}
\usepackage{amssymb}
\usepackage{amsfonts}
\usepackage{graphicx}
\usepackage{dcolumn}
\usepackage{bm}
\usepackage{makecell}
\usepackage{xcolor}
\usepackage{enumerate}
\usepackage[caption=false]{subfig}
\usepackage{float}
\usepackage{csquotes}
\usepackage[unicode=true,pdfusetitle,
 bookmarks=true,bookmarksnumbered=false,bookmarksopen=false,
 breaklinks=false,pdfborder={0 0 1},backref=false,colorlinks=true]
 {hyperref}
\hypersetup{linkcolor=blue,urlcolor=blue,citecolor=blue}

\newcommand{\be}{\begin{equation}}
\newcommand{\ee}{\end{equation}}
\newcommand{\bd}{\begin{displaymath}}
\newcommand{\ed}{\end{displaymath}}
\newcommand{\BE}{\begin{eqnarray}}
\newcommand{\EE}{\end{eqnarray}}

\begin{document}
\title{Finite-size effect in Kuramoto phase oscillators with higher-order interactions}
\author{Ayushi Suman, Sarika Jalan}
\address{Complex Systems Lab, Department of Physics, Indian Institute of Technology Indore, Khandwa Road, Simrol, Indore-453552, India}
\ead{ayushishuman@gmail.com, sarika@iiti.ac.in}
\begin{abstract}
Finite-size systems of Kuramoto model display intricate dynamics, especially in the presence of multi-stability where both coherent and incoherent states coexist. We investigate such scenario in globally coupled populations of Kuramoto phase oscillators with higher-order interactions, and  observe that fluctuations inherent to finite-size systems drives the transition to the synchronized state occurring before the critical point in the thermodynamic limit. Using numerical methods, we plot the first exit time distribution of the magnitude of complex order parameter and obtain numerical transition probabilities across various system sizes. Further, we extend this study to a two-population oscillator system, and using velocity field of the associated order parameters, show the emergence of a new fixed point corresponding to a partially synchronized state arising due to the finite-size effect which is absent in the thermodynamics limit. 
\end{abstract}
\paragraph{\bf{Introduction and Motivation}}
Synchronization, the phenomenon in which coupled oscillators modify their rhythms over time under the influence of each other, has captivated scientists across disciplines for decades. This fascinating behavior arises in a myriad of natural systems (rhythmic flashing of fireflies to the coordinated firing of neurons in the brain \cite{antzoulatos2014increases}) and artificial systems (laser arrays \cite{nixon2013observing}, power grids \cite{motter2013spontaneous}, and large-scale architectures \cite{strogatz2005crowd}). Therefore, understanding the origin of synchronization has profound implications for fields ranging from neuroscience and biology to physics and engineering.
\begin{figure}[b]
\centering
\includegraphics[width=0.5\textwidth]{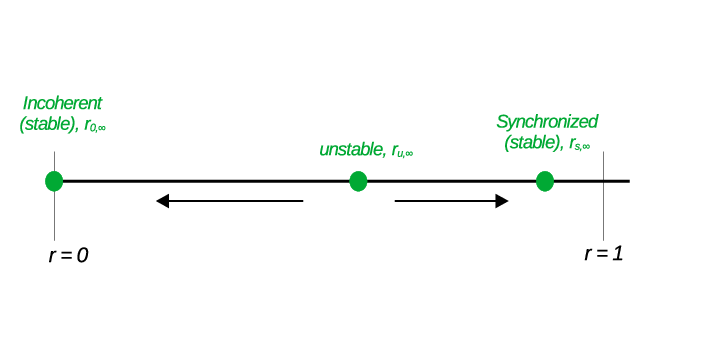}
\caption{Schematic representation of three real, positive solutions of Eq.~\ref{eqn_r} for order parameter measuring the strength of synchronization. One stable incoherent state exists at $r = 0$ (denoted as $r_{0,\infty}$), another stable state exists near $r \sim 1$ (written as $r_{s,\infty}$). One unstable state exists between these two stable states and is denoted as $r_{u,\infty}$.}
\label{fig0_schamatic} 
\end{figure}
The study of synchronization started with the seminal work of Arthur Winfree in the 1960s \cite{winfree1967biological}, which laid the groundwork for understanding how this dynamical state emerges from interactions of limit cycle oscillators. Subsequently, Yoshiki Kuramoto's paper titled `` Self-entrainment of a population of coupled nonlinear oscillators'' \cite{Kuramoto1975InIS} presented the Kuramoto model, a paradigm-shifting framework for studying synchronization. Kuramoto model simplified the analysis of synchronization behaviors from limit cycle oscillators to phase oscillators via phase reduction \cite{leon2019phase,pikovsky2012}, demonstrating that a population of coupled oscillators could synchronize only through phase interactions.

The Kuramoto model stands as a cornerstone in the study of synchronization, providing a versatile framework for investigating the emergence of synchronized states in populations of coupled oscillators 
\cite{acebron2000synchronization,zhang2015explosive,Strogatz_Kura_rev}. Recent research has expanded upon Kuramoto's foundational work, exploring variations of the model to capture additional complexities observed in real world systems. These variations often lead to a synchronization phenomenon, different from the one Kuramoto observed, for instance, explosive synchronization (ES) which has been vastly investigated in the last two decades \cite{sakaguchi1986soluble,pazo2005thermodynamic,acebron2005kuramoto,Skardal_prl2019}. Skardal et. al. \cite{Skardal_NatComm2020} introduced higher-order harmonic coupling terms into the Kuramoto model, enabling the investigation of synchronization phenomena beyond the pair-wise 
interactions. Furthermore, Zhang's work in 2015 introduced adaptation mechanism into the Kuramoto model, allowing oscillators to adjust their interaction strengths based on the collective mean field of the system \cite{zhang2015explosive}. Apart from this, many different synchronization routes and their types have been studied with subtle variations in the Kuramoto model \cite{rajwani2023tiered,jalan2022multiple,gomez2007paths}. These studies supply an elaborate understanding of the macroscopic behaviour of the collective oscillatory modes in the thermodynamic limit.

Daido in the year 1990 noticed fluctuations in the order parameter (a measure of  synchronization of coupled Kuramoto oscillators), induced by finite-size of the underlying system, and published a series of papers about scaling of the critical coupling point for synchronization 
\cite{daido1987scaling,daido1990intrinsic,daido1996onset}. Later, Pikovsky \cite{peter2018transition,komarov2015finite} argued that a transition to synchrony occurs only for finite-size ensembles and disappear in the thermodynamic limit in noise-driven nearly identical oscillators. Strogatz explored finite-size effects in systems with uniformly spaced frequencies, examining the role of frequency distribution on synchronization behavior \cite{ottino2016kuramoto}. Verwoerd and Mason established necessary and sufficient conditions for the existence of fixed points in finite systems of coupled phase oscillators \cite{verwoerd2007conditions,verwoerd2008global}. Additionally, there exist several other notable studies including a field theoretic approach by Buice and Chow \cite{buice2007correlations} addressing various aspects of synchronization in finite-size systems \cite{belykh2015dynamics,tang2011synchronize,bronski2012fully,hong2007finite}. 

Motivated by these empirical observations and theoretical frameworks, our study of finite-size effects extends, first, to higher-order interactions where multistability causes basin jumps, and second, to a two-population oscillator scenario which introduces an additional dimension to the phase space. Accordingly, a central focus of our investigation lies, first, in analyzing the properties of critical coupling at which transition to synchronization occurs due to finite-size effect in presence of higher-order interactions, and second, the emergence of a partially synchronized state, which is realized in two dimensional (2D) phase space (of two-population) due to finite-size effects. A central focus of the previous studies in Kuramoto model has been the shifting of critical coupling and the route to synchronization. We show, using numerical methods, that such a critical point cannot be defined for finite-size systems.

\paragraph{\bf{Model:}}
The model considered for this study includes a pairwise coupling term (as in the Kuramoto model) and one higher harmonic  (triadic nonlinear coupling) as obtained from phase reduction of the Stuart-Landau limit cycle oscillators \cite{leon2019phase}. Inclusion of the higher order coupling offers bistability and facilitates the study of basin jumping as a result of finite-size. Further, we include a mean field adaptation between two identical layers whose coupling strengths are influenced by the extent of synchrony in the other layer. This is motivated by \cite{zhang2015explosive} and extends the (reduced) dimension of the system to a 2D plane where more intricate dynamics can be observed which are absent for dynamics on a line. Reduced dimension refers to the number of dimensions the whole system collapses to after using Ott-Antonsen dimensionality reduction technique. The model equation reads
\begin{equation}
\dot{\theta_i}^{(l)} =\omega_i^{(l)}+g(\boldsymbol{\theta}^{(l')})\Bigg[
\frac{K_1^{(l)}}{N}\sum\limits_{j=1}^{N}\sin(\theta_j^{(l)}-\theta_i^{(l)})
+\frac{K_2^{(l)}}{N^2} \sum\limits_{j,l=1}^{N}\sin(2\theta_j^{(l)}-\theta_l^{(l)}-\theta_i^{(l)})\Bigg],
\label{eq_model}
\end{equation}	
where, \ref{eq_model} $\theta_i$ and $\omega_i$ are, respectively, the instantaneous phase and intrinsic frequency of the $i^{th}$ oscillator with $\omega_i$ sampled from a Lorentzian distribution with mean $\omega_0$ and standard deviation $\Delta$. $K_1$ and $K_2$ are, respectively, the pairwise and triadic coupling strengths. The index $(l)$ and $(l')$ specify the layer index such that $l,l' \in {1,2}$ and $l \neq l'$. The individual layers 1 and 2 will be referred as L1 and L2 respectively. $g$ is an adaptation function which depends on the phases $\theta$ of all oscillators in layer $l'$.

\begin{figure}[ht]
\centering
\includegraphics[width=\textwidth]{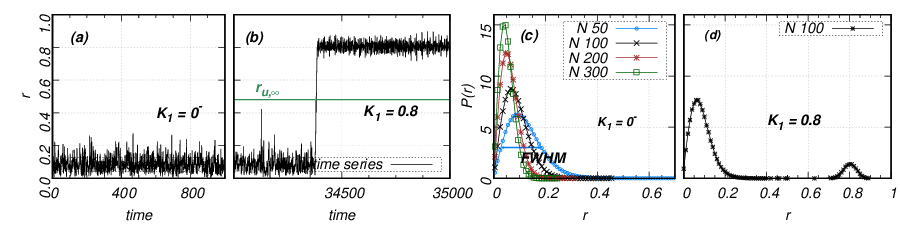}
\caption{(a) time series of $r$ for a small window of time. (b) The time series for $N = 100$ for a larger value of $K_1=0.8$ where synchronization transition is observed in time. (c) Normalized probability distribution obtained from the time series in (a) for $N=100$ is plotted in black. Blue, red and green line with points are the Normalized probability distributions for $N=50,200,300$ respectively for $K_1=0^-$. (d) The normalized probability distribution for the time series in (b). }
\label{fluctuation_and_distribution} 
\end{figure}

To quantify the extent of synchronization, Kuramoto proposed the use of a complex-valued quantity which is equal to the centroid of the instantaneous phases of all the oscillators in the complex plane
\begin{equation}
 z= r^{(l)}e^{\iota \Psi^{(l)}} = \frac{1}{N}\sum\limits_{j=1}^{N} e^{\iota \theta_i^{(l)}},
\label{eq_order_parameter}
\end{equation}
magnitude, $r^{(l)}$, of which provides the fraction of oscillators mutually entrained at mutual frequency $\Psi$. $r^{(l)}\sim 0$ and corresponds to the incoherent state which is observed before the critical point and $r^{(l)} = 1$ corresponds to complete synchrony in the $l^{th}$ layer.  The inter-layer adaptation function, $g(\boldsymbol{\theta^{(l)}})$, in our model is an algebraic power of the magnitude of the order parameter $r^{(l')}$ of the other layer $(l')$ such that $g(\boldsymbol{\theta}^{(1)}) = r^\alpha$ and $g(\boldsymbol{\theta}^{(2)}) = \rho^\alpha$. 
\paragraph{\bf{Single layer case:}}
For first part of the study, we restrict $\alpha = 0$ which decouples both layers from each other. Thus, the layer index $l$ and $l'$ becomes redundant and the model reduces to
\begin{equation}
\dot{\theta_i} =\omega_i+\frac{K_1}{N}\sum\limits_{j=1}^{N}\sin(\theta_j-\theta_i)
+\frac{K_2}{N^2} \sum\limits_{j,l=1}^{N}\sin(2\theta_j-\theta_l-\theta_i).
\label{model_1d}
\end{equation}
Using the fact that the number of oscillators is conserved,  Ott and Antonsen \cite{Ott_Antenson2008} in 2008 suggested an ansatz which, in the thermodynamic limit ($N \to \infty$) collapses the system to a 1D manifold (refer supplementary material for complete derivation) and the temporal evolution of the system is led by a nonlinear differential equation in $r$ which is the order parameter in $N \to \infty$ limit,
\begin{equation}
\dot{r} = f(r) \\
=-r \Delta + \frac{K_1}{2}(r - r^3) + \frac{K_2}{2}(r^3 - r^5),
\label{eqn_r}
\end{equation}
such that $f(r_\infty) = 0$ yields all the fixed points of $r$. This equation admits three real, positive solutions between the backward critical point $K_{c,b,\infty}$ and the forward critical point $K_{c,f,\infty}$ \cite{Skardal_NatComm2020} (Fig.~\ref{fig0_schamatic}). $r_\infty = 0 = r_{0,\infty}$ is a trivial case and corresponds to the incoherent state. $0<< r <1$ corresponds to a stable point where a fraction $r$ of oscillators are phase locked and this state is denoted as $r_{s,\infty}$. The third solution corresponds to an unstable state which acts as a basin boundary between the incoherent and the synchronized state. This state is denoted as $r_{u,\infty}$.
The unstable point pushes the trajectory of $r$ away from it while the stable point attracts the trajectory towards it. Thus, in $t\to\infty$, the system always settles at a stable fixed point unless started at $r_{u,\infty}$ (Fig.~\ref{fig0_schamatic}). For a finite-size system, the stationary states of $r$ do not coincide with $r_\infty$ but fluctuate strongly depending upon the system size (Fig.~\ref{fluctuation_and_distribution}(a)) with relative fluctuation $\frac{\Delta r}{\langle r \rangle}  \sim \mathcal{O}(1/\sqrt{N})$ where $\Delta r$ and $\langle r \rangle$ are, respectively,  the variance and mean of the time series $r$ forms. A dynamical system on a 1D line is always a gradient system which means there exists a continuous and differentiable scalar function $V(r)$ such that an over-damped Brownian particle in this potential $V$ can give a complete understanding of the quantity $r$ in finite-size Kuramoto model. Unfortunately, there is no known solution to such a nonlinear potential and we resort to numerical methods.

We simulate a system with $N$ oscillators for various N and evolve them using Eq.~\ref{model_1d} with $rk-4$ method for $t \sim 10^5$ time steps. We take two cases, $K_1 < 0$ and $K_1 > 0$ because $0$ is the bifurcation point for $K_2=8$. Note that, unless stated otherwise, $K_2 = 8$ is the value taken in this entire study (for other values of $K_2$, refer supplementary material and \cite{Skardal_NatComm2020} ). In the first case, only the incoherent state is accessible to system while in the second case, both incoherent and synchronized states are accessible. The time series for $r$ for $K_1 = -0.1$ is plotted in Fig.~\ref{fluctuation_and_distribution}(a) for a small interval of time in the incoherent regime i.e. $K_1 < K_{c,b,\infty}$. The intrinsic frequency distribution is sampled uniformly from a Cauchy distribution with mean $\omega_0 = 0$ and standard deviation $\Delta = 1$. The intrinsic frequency distribution and the sampling style has a significant effect on synchronization and is studied in \cite{peter2018transition}. In $t\to\infty$, $r$ forms a time dependent probability distribution $P(r)$ which is depicted with black cross points in Fig.~\ref{fluctuation_and_distribution}(c), and is obtained by normalizing a histogram formed by the data in Fig.~\ref{fluctuation_and_distribution}(a) over 50 realizations. Blue open circle, black cross, red star and green open square points give the probability distributions, respectively, for $N=50, 100, 200$ and $300$. It can be observed from  Fig.~\ref{fluctuation_and_distribution}(c) that a smaller size system has a larger FWHM (full width at half maximum). Due to the fluctuations, the trajectory in incoherent state has a finite non-zero probability of reaching $r_{u,\infty}$ and thus crossing the basin boundary to reach $r_{s,\infty}$ where the trajectory then keeps fluctuating around $r_{s,\infty}$ \ref{fluctuation_and_distribution}(b) such that $\langle r \rangle_{\Delta t \to\infty} = r_{s,\infty}$. To demonstrate this, we take $K_1 = 0.8$, where such a transition is observed in computationally realizable number of time steps. The probability distribution of $r$ over 50 realizations for $K_2=0.8$ is plotted in Fig.~\ref{fluctuation_and_distribution}(d) which is observed to be a bimodal distribution (one mode corresponds to incoherent state and the other corresponds to synchronized state).

\begin{figure*}[ht]
\centering
\includegraphics[width=0.8\textwidth]{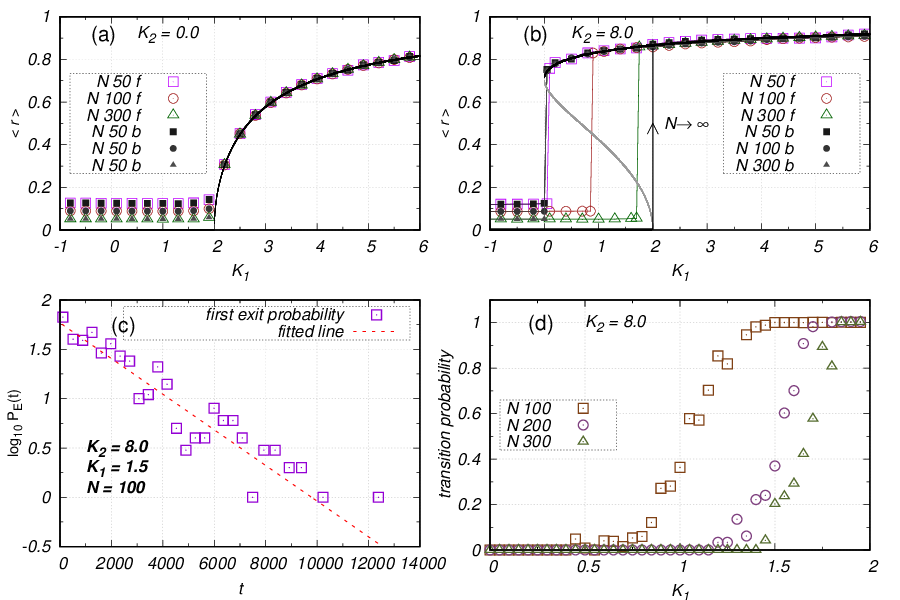}
\caption{Dependence of critical coupling strength on network size. The black, dark grey and light grey curves denote the stable, unstable and saddle points, respectively with variation in $K_1$. This is the solution of dimensional reduced system in thermodynamic limit (Eq.~\eqref{eqn_r}).  Magenta open squares, brown open circles and green open triangles correspond to numerical simulation of the finite size system for sizes $50, 100 $ and $300$, respectively. The postscript $b$ and $f$ specify the direction, backward and forward, respectively, for adiabatic change in $K_1$. (a) For $K_2 = 0$, (b)  for $K_2 = 8$. (c) The exit time distribution on a semi log graph for $K_2 = 8$, $K_1 = 1.5$, and $N = 100$. Open square are data points obtained from simulation and red dotted line is a straight line fit. (d) Probability of synchronization with $K_1$ for different $N$ values. The brown open square, purple open circle and green open triangle points are obtained by dividing the number of realizations where system synchronizes by the total number of realizations.}
\label{sync_profile_fit_and_probability} 
\end{figure*}

A system with higher FWHM has a higher probability of reaching $r_{u,\infty}$ and thus jumping to the synchronized basin. Therefore, the smaller is the size of a system, it is more probable to synchronize. Fig.~\ref{sync_profile_fit_and_probability}(a) and  \ref{sync_profile_fit_and_probability}(b) show the variation of mean order parameter ($ \langle r \rangle_{\Delta t \to\infty}$) with coupling strength $K_1$ for $K_2=0$ and  $K_2=8$, respectively. This is simulated with an initial condition $\boldsymbol{\theta}$ and evolved for $10^5$ time steps across the $K_1$ axis in both forward and backward directions. For forward direction, $\boldsymbol{\theta}$ is distributed uniformly on the unit circle and for backward direction, $\boldsymbol{\theta} = 0$.
$K_2=0$ is a case where multistability does not exist and the critical point of transition is approximately the same as $K_{c,\infty}$ at the scale in consideration. The stable and unstable fixed points for $N\to\infty$ are plotted as black and grey lines, respectively (Fig.~\ref{sync_profile_fit_and_probability}(a) and \ref{sync_profile_fit_and_probability}(b)). These are obtained by solving $f(r_\infty)=0$. Magenta open square, red open circle and green open triangle correspond to time averaged $\langle r \rangle$ for $N=50,100$ and $300$, respectively in the forward direction adiabatically and black filled squares, circles, and triangles show the $\langle r \rangle$ for backward direction. In Fig.~\ref{sync_profile_fit_and_probability}(b), the de-synchronization point is same for the range of $N$ considered while the forward transition point changes with $N$. The system sizes are chosen such that the forward synchronization transition spans the whole bistable region. For larger $N$, the forward synchronization transition saturates near $K_1 = 2$. For $K_2=8$, due to the bistability,  basin jumping comes into play thereby yielding a shift into the forward critical point of synchronization  towards left (Fig.~\ref{sync_profile_fit_and_probability}(b)). This shift is larger for smaller $N$. The critical point of synchronization in these simulations is not well defined but varies with different realizations of the simulation. Between $K_{c,b,\infty}$ and $K_{c,f,\infty}$, there is a non zero probability of synchronization and this probability increases with $K_1$ because the unstable point keeps coming closer to $r = 0$ as $K_1$ is increased.

We now use the concept of first passage time\cite{redner2023first} from random walker, which describes the time it takes for a stochastic process to leave a certain region or state for the first time. In this system of Kuramoto oscillators, if $P_E(t)$ is the exit time distribution ,i.e. the distribution of time it takes for a system to leave the incoherent state for the first time, we obtain an exit time distribution using an initial condition where $\boldsymbol{\theta}$ is distributed randomly on a circle thus setting $r=0$. The system evolves until it crosses the basin boundary and escapes the meta stable state, the initial condition is again reset. Using $10000$ reset data, we plot the exit time distribution on a semi log graph which is a straight line Fig.~\ref{sync_profile_fit_and_probability}(c). The unnormalized distribution fits
\begin{equation}
P_E(t)=Ae^{-\lambda t},
\label{exponential_distribution}
\end{equation}
where $A$ is a constant and $\lambda$ is the rate parameter, which is also called the hazard rate. For $K_1=1.5$, $K_2=8$ and $N=100$, a straight line fit on semi log graph gives $A=60.062$ and $\lambda=4.1 \times 10^{-4}$. The mean exit time $\mathcal{E}[T]$ is the reciprocal of $\lambda$, $\lambda=\frac{1}{\mathcal{E}[T]}$. The rate parameter $\lambda$ gives the probability per unit time of system leaving the incoherent state and $\int_0^t Ae^{-\lambda t}$ gives the probability of synchronization in time $0$ to $t$. Thus the survival function will be $S_E(T) = P(T>t)=\int_t^\infty A e^{-\lambda t}$. When the exit time distribution is exponential, it suggests that the system has a memory-less property. i.e. the probability of leaving the inherent state is independent of how long the system has already been in this state. Hence an ensemble averaging gives a numerical probability of transition to the synchronized state. This is calculated by taking $50$ realizations and counting the number of times system transitions to synchronization. This \emph{transition probability} is plotted in Fig.~\ref{sync_profile_fit_and_probability}(d) for $N=100,200$ and $300$.
\begin{figure}[t]
\centering
\includegraphics[width=0.9\columnwidth]{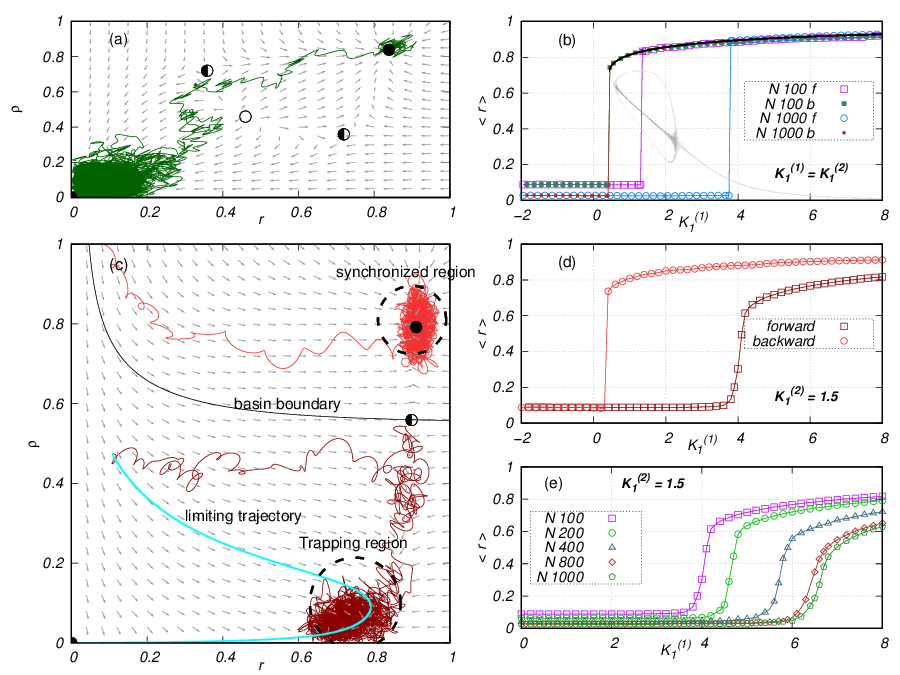}
\caption{ (a) Trajectory of the macroscopic system in $r-\rho$ plane is plotted in green and the vector field in plotted in grey. This vector field gives the direction of $(\Dot{r},\Dot{\rho})$ in $r-\rho$ plane. Open, filled and half filled circles represent the unstable, stable, and saddle points for $K_1^{(1)} = K_1^{(2)}=1.5$. (b) Synchronization transition for $N=100$ is shown by magenta open squares for forward direction and green closed square in backward direction. For $N=1000$, sky-blue open circles for forward direction and red closed circle for backward direction. (c) At $K_1^{(2)}=1.0$, $K_1^{(1)}=6.0$, $\alpha = 0.3$ and $N = 100$, two trajectories are shown in the phase space. Light-red trajectory originates from $(0.1,1)$ goes to synchronized state and dark-red trajectory originating close to $(0.1,0.5)$ goes to the trapping region. Cyan thick line shows the time evolution for same initial condition in infinite limit. (d) Forward and backward transition. Both the stationary states from (a) are shown on $K_1^{(1)}$ axis in the same colour code. (c)The forward transitions for $N = 100, 200, 400, 800$ and $ 1000$ is plotted in magenta open square, light-green open circle, blue open triangle, brown open diamond, and dark-green open pentagon respectively. }
\label{special_case} 
\end{figure}
\paragraph{\bf{Multilayer case:}}
We next extend the study of $finite-N$ to a two interacting population system. $g\boldsymbol{(\theta^{(l)})}$ for L1, as specified earlier, is taken to be a real algebraic power of the magnitude of order parameter of L1 and similarly for L2, $ g\boldsymbol{(\theta^{(l)})} = \Bigg|\frac{1}{N} \sum\limits_{j=1}^{N} e^{\iota \theta_i^{(l)}}\Bigg| ^\alpha$ where $l,l' \in {1,2}$ with $l \neq l'$. Writing 
 $s = 1/N \sum\limits_{j=1}^{N} e^{\iota \theta_i^{(2)}}$, the complex order parameter for L2. such that $\rho=\Bigg|\frac{1}{N} \sum\limits_{j=1}^{N} e^{\iota \theta_i^{(l)}}\Bigg|$, the corresponding dimensionally reduced coupled differential equations are
\begin{equation}
\begin{split}
\dot{r}
=-r \Delta + \bigg(\frac{K_1}{2}(r - r^3) + \frac{K_2}{2}(r^3 - r^5)\bigg)\rho^\alpha ,\\
\dot{\rho} 
=-\rho \Delta + \bigg(\frac{K_1}{2}(\rho - \rho^3) + \frac{K_2}{2}(\rho^3 - \rho^5)\bigg) r^\alpha .
\end{split}
\label{eqn_r rho}
\end{equation}
These coupled equations \emph{(refer supplementary material for derivation)} provide the time evolution for any initial condition $(r,\rho)$ and the dynamics is now confined on a 2D plane. On solving Eq.~\ref{eqn_r rho} computationally, we find a maximum of two stable points for any parameter values. $(0,0)$, which is locally stable for all couplings, corresponds to the incoherent state, whereas $(r_{s,\infty},\rho_{s,\infty})$ corresponds to the synchronized state. Similar to the previous case, the trajectory of system now fluctuates in a 2D plane with a drift in the $(\Dot{r},\Dot{\rho})$ direction. A velocity field is plotted in grey in Fig.~\ref{special_case}(a) with arrows indicating the direction of drift for $K_1^{(1)} = K_1^{(2)}$. Black open, filled and half filled circle shows the unstable, stable and saddle points $K_1^{(1)} = K_1^{(2)}=1.5$. The green line shows the trajectory of system in $(r,\rho)$ plane. Fig.~\ref{special_case}(b) shows the synchronization and de-synchronization transition for $N=100$ and $1000$ in both forward and backward direction. Black, dark-grey and light-grey lines show the stable, unstable and saddle points with $K_1^{(1)}$ in Fig.~\ref{special_case}(b). 

If one of the pairwise coupling strength, say $K_1^{(2)}$, is small ($1.5$ in Fig.~\ref{special_case}(c)), then L1 will undergo a second order transition as a result of finite size in the coupled system\emph{(refer supplementary material for higher values for $K_1^{(2)}$)}. Fig. \ref{special_case}(c) shows a vector field which gives the direction of velocity(phase flow) in $(r,\rho) \in (0,1)$ plane. Black solid line in the basin boundary which separates the incoherent and synchronized basins. If we evolve any initial condition in this vector field in the $N\to\infty$ limit, the accessible states are $(0,0)$ and $(r^*_{\infty, s},\rho^*_{\infty, s})$ depending upon the basins of attraction. For smaller values of $\alpha$, due to finite size, $(0,0)$ becomes inaccessible because the trajectory occupies a certain non zero volume due to fluctuations in the stationary state. In Fig.~\ref{special_case}(c), (d) and (e), $\alpha$ is fixed to $0.3$, vector field for other values of $\alpha$ are shown in supplementary material. The stationary states are not points, but can be thought of as finite, non-zero volume objects in the phase space. Two trajectories, for two initial conditions, are plotted in Fig.~\ref{special_case}(c) in light-red and dark-red for synchronized and incoherent basins respectively. The phase flow near $\rho=0$ axis is in opposite direction to that at $\rho = 0$ and the finite $N$ trajectory experiences a torque in two opposite direction and a trapping region is created. This trapping region changes it's position based on system parameters and is plotted in Fig.~\ref{special_case}(d). The forward (dark-red open squares) and backward (light-red open circles) transitions with $K_1^{(1)}$ is shown in Fig.~\ref{special_case}. Thus, L1 has both forward and backward transitions without a hysteresis. Fixed point thus created due to finite size of L1 and L2 do not correspond to any attracting point in infinite limit. Due to this trapping region, a second order transition can be observed despite the presence of higher-order interaction.
The synchronization transition (second order) for various $N$ is plotted in Fig.~\ref{special_case}(c). A smaller $N$ creates a larger volume of stationary state and therefore synchronizes earlier than that of a large $N$ system while in the infinite limit, forward synchronization disappears altogether.

\paragraph{\bf{Conclusion:}}
Fluctuations due to finite size in an ensemble of Kuramoto oscillators can lead to a phase transition when the system inherently does not have a critical point in the thermodynamic limit. The effective critical point of the transition in such systems cannot be exactly predicted.
Present paper shows that phase transition to synchronization can occur anywhere between a certain range of parameter (pairwise coupling strength) values with some probability which we obtain numerically. We consider the behaviour of magnitude of order parameter as a stochastic process and plot a first exit time probability. Then using the Ott-Antonsen dimensionality reduction method, we obtain an easy to compute relation that yields all the fixed points in the thermodynamic ($N \to \infty$) limit. We then hypothesize that a $finite-N$ system forms stationary states where the macroscopic order parameter keeps fluctuating around the fixed points and its time averaged value equals the fixed point in $N \to \infty$. In the Kuramoto model with higher-order interactions, there exists two bifurcations on the pairwise coupling scale; a saddle node bifurcation creates a synchronized state along with an unstable state which acts as the basin boundary between the incoherent and the synchronized state, and a sub-critical pitchfork bifurcation where the previously unstable branch merges with the fixed point for incoherent state. Between these two points, due to multistability, the order parameter has a finite non zero probability of jumping from incoherent basin to the synchronized basin. This probability of transition increases with an increase in the pairwise coupling as the distance between the fixed point (incoherent state) and the basin boundary decreases while the relative fluctuation remains the same. 

Also, by considering a two population system (which reduces to 2D phase plane for macroscopic order parameters), we show that  stationary states can be formed away from the fixed points in $N \to \infty$. We demonstrate this by using the vector field of the order parameters of two populations. What follows that this setup leads to a second order phase transition despite the presence of higher order coupling which was  previously known to cause a first order phase transition.

As all the real world interacting systems are finite in size, a study of $finite-N$ in the celebrated Kuramoto model provides hope for a better understanding in applicable areas. In our simulations, we have considered a global coupling while the effect of a network structure also calls for extensive investigation. 

\section*{Acknowledgement}
SJ gratefully acknowledges SERB Power grant SPF/2021/000136 and A. Suman gratefully acknowledges CSIR for funding this research under award no. CSIRAWARD/JRF-NET2021/11014. The authors acknowledge Prashant Gade for his discussions on exit probability. The authors are also thankful to Bapan Ghosh, and the friends at Complex Systems Lab for fruitful discussions. 

\section*{Refrences}

\end{document}


\maketitle
\section{Dimensional reduction}
Writing model Eq. 1 from the main text as
\begin{equation}
    \label{predict}
	\dot{\theta_i}^{(l)} =\omega_i^{(l)}+ g(\boldsymbol{\theta}^{(l^\prime)})[ \hat{H}_i^{(l)} ]
\tag{A}
	\end{equation}
where, 
\begin{eqnarray*}
\hat{H}_i^{(l)} =\frac{K_1^{(l)}}{N}\sum\limits_{j=1}^{N}\sin(\theta_j^{(l)}-\theta_i^{(l)}) 
 + \frac{K_2^{(l)}}{N^2} \sum\limits_{j=1}^{N}\sum\limits_{l=1}^{N}\sin(2\theta_j^{(l)}-\theta_l^{(l)}-\theta_i^{(l)})  \nonumber\\
\end{eqnarray*}
Here, the superscript $l$ ( $l^\prime$), taking values 1 and 2,  denote the index of the layer in the consideration, and $K_i^{(l)}$ indicates  the  overall coupling strength for the $i$-simplex  interaction of the $l^{th}$ layer.  In the respective mean fields, the dynamical evolution equations for the simplicial layers can be written as
\begin{equation*}
	\dot{\theta_i}^{(l)} =\omega_i^{(l)}+g(\boldsymbol{\theta}^{(l^\prime)})\bigg[K_1^{(l)}r_1^{(l)}\sin(\Psi_1^{(l)}-\theta_i^{(l)}) + K_2^{(l)}r_1^{(l)}r_2^{(l)}\sin(\Psi_2^{(l)}-\Psi_1^{(l)}-\theta_i^{(l)}) \bigg]\nonumber \\,
	\label{eq_mean} 
        \tag{B}
	\end{equation*}
with the complex order parameters of the $l^{th}$ simplicial layer defined as, 
\begin{equation*}
 z_q^{(l)}= 
r_q^{(l}e^{\iota\Psi_q^{(l)}}=\frac{1}{N}\sum\limits_{j=1}^{N}e^{\iota q\theta_j^{(l)}}
\label{eq_order}
\end{equation*}

Using the complex order parameters (\ref{eq_order}) and defining $ g\boldsymbol{(\theta^{(l)})} = \Bigg|\frac{1}{N} \sum\limits_{j=1}^{N} e^{\iota \theta_i^{(l)}}\Bigg| ^\alpha$ where $l,l' \in {1,2}$ with $l \neq l'$ and $\alpha \in (0,1)$, Eq.~\ref{eq_mean} can be written as 
\begin{equation}
 \dot{\theta_i}^{(l)}=\omega_i^{(l)}+\frac{r_1^{(l^\prime)\alpha}}{N}(H^{(l)}e^{-\iota\theta_i^{(l)}}-H^{*(l)}e^{\iota\theta_i^{(l)}})  
 \label{eq_mean2}
 \tag{C}
\end{equation}
 with $H^{(l)}=K_1^{(l)} z_1^{(l)}+K_2^{(l)} z_2^{(l)} z_1^{*(l)}$.
 In the thermodynamic limit $N \longrightarrow \infty$, the state of the individual layer can be described by a density function $f^{(l)}(\theta^{(l)},\omega^{(l)},t)$ which measures the density of oscillators with phase between $\theta^{(l)}$ and $\theta^{(l)} + d\theta^{(l)}$ having natural frequency lying between $\omega^{(l)}$ and $\omega^{(l)} + d\omega^{(l)}$ at time $t$ for the $l^{th}$ layer. 
 Since the number of oscillators in each layer is conserved, the density functions will individually satisfy the continuity equation,
{\small{
\begin{equation}
    0=\frac{\partial {f^{(l)}}}{\partial t}+\frac{\partial}{\partial \theta^{(l)}} \left[{f^{(l)}} \left[ \omega_i^{(l)} + \frac{r_1^{(l^\prime)\alpha}}{N} (H^{(l)}e^{-\iota\theta^{(l)}}-H^{*(l)}e^{\iota\theta^{(l)}}) \right] \right]
    \label{eq_continuity}
    \tag{D}
\end{equation}}}
Assuming the natural frequency $\omega^{(k)}$ of each oscillator to be drawn from a distribution $g(\omega^{(k)})$, the density function $f^{(l)}$ can be expanded into Fourier series as
{\small{
\begin{equation}
    f^{(l)}(\theta^{(l)},\omega^{(l)},t) 
    =\frac{g(\omega^{(l)})}{2\pi} \left[1+\sum\limits_{n=1}^{\infty}\hat{f_n}^{(l)}(\omega^{(l)},t)e^{\iota n\theta^{(l)}} + c.c. \right],\nonumber
    \end{equation}}}
where $\hat{f_n}^{(l)}(\omega^{(l)},t)$ is the $n^{th}$ Fourier component and c.c. are the complex conjugates of the former terms. Next, we use the Ott-Antonsen [ref. 29 main text] ansatz which assumes that all the Fourier modes decay geometrically, i.e., $\hat{f_n}^{(l)}(\omega^{(l)},t)=\alpha^{(l)n}(\omega^{(l)},t)$ for some function $\alpha^{(l)}$ which is analytic in the complex $\omega^{(l)}$ plane. After inserting the Fourier expansion of $f^{(l)}(\theta^{(l)},\omega^{(l)},t)$ in the continuity equation (\ref{eq_continuity}), the dynamics of the two-layer network collapses into a complex two dimensional manifold (Ott-Antonsen manifold).
\begin{equation}
    \dot{\alpha}^{(l)}=-\iota \omega^{(l)}\alpha^{(l)}+\frac{r_1^{(l^\prime)\alpha}}{2} \left[H^{(l)*}-H^{(l)}\alpha^{(l)2}  \right]
    \label{eq_alpha}
    \tag{E}
\end{equation} 
with $H^{(l)}$ defined in Eq.~\ref{eq_mean2}. 
The order parameter in the thermodynamic limit can then be given as\\ $z^{(k)}$=$\int \int f^{(l)}(\theta^{(l)},\omega^{(l)},t)e^{\iota\theta^{(l)}}d\theta^{(l)} d\omega^{(l)}$, which after inserting the Fourier decomposition of $f^{(l)}$ becomes,
\begin{equation}
    z^{(l)}=\int \alpha^{(l)}(\omega^{(l)},t)g(\omega^{(l)})d\omega^{(l)}
    \nonumber
    \end{equation}
    If we choose $g(\omega)$ to be a Lorentzian frequency distribution $g(\omega)=\frac{\Delta}{\pi\left[\Delta^{2}+(\omega-\omega_0)^2\right]}$, where $\omega_0$ is mean and $\Delta$ is the standard deviation. $z^{(l)*}$ can be calculated by contour integration in the negative half complex plane, yielding, $z^{(l)*}=\alpha^{(l)}(\omega_0^{(l)}-\iota\Delta^{(l)},t)$.
For simplicity of notation, we redefine the order parameters as $z^{(1)}=re^{\iota\Phi}$ and $z^{(2)}=\rho e^{\iota\chi}$, $\Phi$ and $\chi$ being the mean phases of layers 1 and 2 respectively.  

\begin{eqnarray*}
        \dot{r}&=&-\Delta r+ \rho^\alpha \bigg[  \frac{K_1^{(1)}}{2}r(1-r^2)+\frac{K^{(1)}_{2}}{2} r^3(1-r^2)\bigg]\nonumber \\
    \dot{\rho}&=&-\Delta\rho+ r^\alpha \bigg[ \frac{K^{(2)}_1}{2}\rho(1-\rho^2)+\frac{K^{(2)}_{2}}{2} \rho^3(1-\rho^2)\bigg]\nonumber \\
    \dot{\Psi}&=&\omega_0^{(1)}  \nonumber \\
    \dot{\chi}&=&\omega_0^{(2)} \nonumber \\
\end{eqnarray*}
For $\alpha=0$, $r$ and $\rho$ uncouple and the superscript $(l)$ becomes redundant giving Eq. 4 (main text).
\newpage
\section{Bifurcation diagrams for $\alpha  \in [0,1]$}
\begin{figure}[H]
\includegraphics[width=\columnwidth]{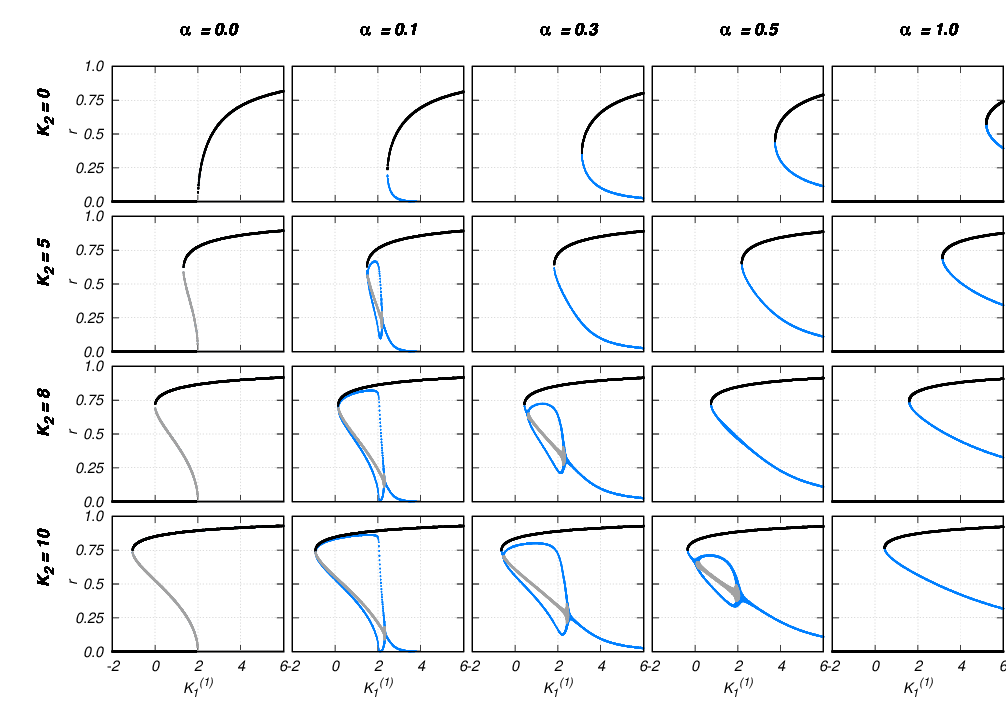}
\caption{Bifurcation diagrams for some values of $\alpha$ and $K_2$. Stable, unstable and saddle points are plotted in black, grey and sky blue respectively. }
\label{many_alpha_k} 
\end{figure}

\section{Simulations for higher values of $K_2$}
\begin{figure}[H]
\includegraphics[width=\columnwidth]{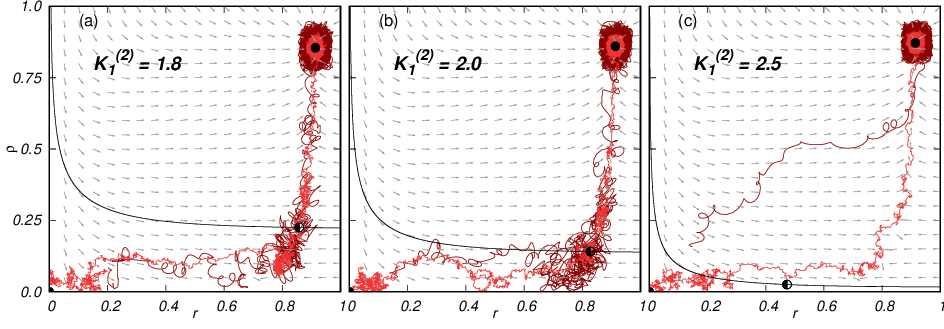}
\caption{Simulation results for trajectories in $(r,\rho)$ plane and the corresponding vector field in $N \to \infty$ limit. $N=100$ is plotted in dark-red line and $N = 500$ is plotted in light-red for (a) $K_1^{(2)} = 0.8$, (b) $K_1^{(2)} = 2.0$, and (c) $K_1^{(2)} = 2.5$,}
\label{k2_higher} 
\end{figure}